\documentclass[twoside,twocolumn,english,prl,showpacs]{revtex4}
\usepackage{ae}
\usepackage{aecompl}
\usepackage[T1]{fontenc}
\usepackage[latin1]{inputenc}
\usepackage{amsmath}
\usepackage{graphicx}
\usepackage{amssymb}

\makeatletter
\newcommand\C[1]{}

\usepackage{babel}
\makeatother
\begin{document}
\newcommand{\Tc}{T_{\mathrm{c}}}
\newcommand{\Lp}{L_{\parallel}}
\newcommand{\Ls}{L_{\perp}}
\newcommand{\fex}{f_{\mathrm{ex}}}
\newcommand{\Fc}{F_{\mathrm{Cas}}}
\newcommand{\Fi}{F_{\mathrm{int}}}
\newcommand{\Xc}{\vartheta}

\title{The Thermodynamic Casimir Effect in $^{4}$He Films near $T_{\lambda}$:
Monte Carlo Results}

\author{Alfred Hucht}

\affiliation{Theoretical Physics, University of Duisburg-Essen, 47048 Duisburg,
Germany}

\email{fred<at>thp.Uni-DuE.de}

\begin{abstract}
The universal finite-size scaling function of the critical Casimir
force for the three dimensional XY universality class with Dirichlet
boundary conditions is determined using Monte Carlo simulations. The
results are in excellent agreement with recent experiments on $^{4}$He
Films at the superfluid transition and with available theoretical
predictions.
\end{abstract}

\date{September 12, 2007}

\pacs{68.35.Rh, 05.10.Ln, 05.50.+q, 64.60.Fr}

\maketitle
Casimir forces are always present in nature when a medium with long-range
fluctuations is confined to restricted geometries. The quantum mechanical
Casimir effect was proposed theoretically 60 years ago by H.~B.~G.~Casimir
\cite{Casimir48} and describes an attractive force between two conducting
plates in vacuum, induced by the vacuum fluctuations of the electromagnetic
field. Furthermore, Goldstone modes \cite{LiKardar91} and surface
fluctuations \cite{ZandiRudnickKardar04} can give rise to Casimir
forces. Near continuous phase transitions, long-range fluctuations
of the order parameter lead to the analogous \emph{thermodynamic}
Casimir effect \emph{}\cite{FisherdeGennes78}, which can change the
thickness of critical liquid films \cite{Indekeu86a}. In a series
of papers, Garcia and Chan \cite{GarciaChan99url,GarciaChan02} and
Ganshin \emph{et al.}~\cite{GanshinScheidemantelGarciaChan06} were
able to measure the thinning of liquid $^{4}$He films close to the
$\lambda$-point due to the thermodynamic Casimir effect. They found
a characteristic deep minimum (dip) in the film thickness just below
the superfluid transition temperature $T_{\lambda}$. Using finite-size
scaling methods, they accurately determined the scaling function $\Xc(x)$
of the Casimir force, which is universal for given universality class
and boundary conditions. For liquid $^{4}$He films it is believed
that the superfluid order parameter vanishes at both surfaces of the
film, implying Dirichlet boundary conditions \cite{HuhnDohm88}.

Unfortunately, a theoretical explanation of the strong dip and a determination
of the scaling function $\Xc(x)$ is still lacking. In Ref.~\cite{DantchevKrechDietrich05},
this is stressed as the main theoretical problem with respect to the
explanation of the $^{4}$He experiments. While field theoretical
results \cite{KrechDietrich92a,KrechDietrich92c,DiehlGruenebergShpot06}
are restricted to temperatures $T\ge\Tc$, Monte Carlo simulations
are only available for periodic boundary conditions until now \cite{DantchevKrech04},
as only in this case the used stress tensor representation of the
Casimir force is applicable. Recent attempts \cite{ZandiShackellRudnickKardarChayes07,MaciolekGambassiDietrich07}
to explain the strong dip within mean field theories only find qualitative
agreement with the experiments, neglecting the non-critical contributions
of Goldstone modes.

In this letter, I present a direct calculation of the Casimir force
using Monte Carlo simulations of the classical XY model. This method
requires the computation of the free energy of the system with high
accuracy, which is a major challenge within Monte Carlo simulations.
Fortunately, it turns out that the determination is greatly simplified
by the fact that only the difference of two free energies is needed,
which goes to zero exponentially fast above $\Tc$. The resulting
finite-size scaling function is in excellent agreement with the experimental
results \cite{GarciaChan99url,GarciaChan02,GanshinScheidemantelGarciaChan06}
as well as with available theoretical predictions \cite{Indekeu86,KrechDietrich92a,LiKardar91}. 

The Casimir force per unit area of a system with size $\Lp^{d-1}\times\Ls$
($\Lp\rightarrow\infty$) is defined as \cite{BrankovDantchevTonchev00,Krech94}
\begin{equation}
\beta\Fc(T,\Ls)=-\frac{\partial\fex(T,\Ls)}{\partial\Ls},\label{eq:betaF_Casimir}\end{equation}
where $\fex(T,\Ls)$ denotes the excess free energy\begin{equation}
\fex(T,\Ls)=f(T,\Ls)-\Ls f_{\infty}(T)\label{eq:f_ex}\end{equation}
of the system. Here $f(T,\Ls)$ is the free energy per unit area of
a film of thickness $\Ls$, measured in units of $k_{\mathrm{B}}T$,
and $f_{\infty}(T)$ is the bulk free energy density. For large $\Ls$
and near $\Tc$ the Casimir force fulfills the scaling \emph{ansatz}
\footnote{$\sim$ means {}``asymptotically equal'' in the respective limit,
$\Ls\rightarrow\infty$, $T\rightarrow\Tc$, or both, keeping the
scaling variable $x$ fixed, e.g. $f(L)\sim g(L)\Leftrightarrow\lim_{L\rightarrow\infty}f(L)/g(L)=1$ %
} \begin{equation}
\beta\Fc(T,\Ls)\sim\Ls^{-d}\Xc(x)\label{eq:X_Casimir}\end{equation}
with the universal finite-size scaling function $\Xc(x)$ and the
scaling variable $ $\begin{equation}
x=t\left(\frac{\Ls}{\xi_{0}^{+}}\right)^{1/\nu}\overset{t>0}{\sim}\left(\frac{\Ls}{\xi_{\infty}^{+}(t)}\right)^{1/\nu}.\label{eq:x}\end{equation}
Here I introduced the reduced temperature $t=T/\Tc-1$ and the bulk
correlation length \begin{equation}
\xi_{\infty}^{+}(t)\sim\xi_{0}^{+}t^{-\nu},\quad(t>0).\label{eq:xi}\end{equation}
Note that the universal finite-size scaling function $\Xc(x)$ depends
on the boundary conditions in the $\Ls$-direction.

We can directly calculate the Casimir force (\ref{eq:betaF_Casimir})
by integration of the internal energy as follows: Let us define the
{}``internal Casimir force'' \begin{equation}
\beta\Fi(T,\Ls)=-\left(\frac{\partial u(T,\Ls)}{\partial\Ls}-u_{\infty}(T)\right)\label{eq:betaF_internal}\end{equation}
with the internal energy per unit area in units of $k_{\mathrm{B}}T$
\begin{equation}
u(T,\Ls)=-T\frac{\partial f(T,\Ls)}{\partial T}\label{eq:u}\end{equation}
and the corresponding bulk density $u_{\infty}(T)$. The quantity
$\partial u(T,\Ls)/\partial\Ls$ is directly accessible in Monte Carlo
simulations using $u=\langle\beta\mathcal{H}\rangle/\Lp^{2}$, and
the central difference quotient\begin{equation}
\frac{\partial u(T,\Ls)}{\partial\Ls}\approx\frac{u(T,\Ls+1)-u(T,\Ls-1)}{2}.\label{eq:differenceQuotient}\end{equation}
The Casimir force is then obtained by integration \begin{equation}
\beta\Fc(T,\Ls)=-\int_{T}^{\infty}\frac{\mathrm{d}\tau}{\tau}\beta\Fi(\tau,\Ls).\label{eq:Integral}\end{equation}
By Eqs.~(\ref{eq:X_Casimir}) and (\ref{eq:Integral}), the internal
Casimir force fulfills the scaling form\begin{equation}
-\beta\Fi(T,\Ls)\sim(\xi_{0}^{+})^{-1/\nu}\Ls^{(\alpha-1)/\nu}\Xc'(x)\label{eq:X_internal}\end{equation}
with the universal finite-size scaling function $\Xc'(x)$. Note that
within the scaling regime, Eq.~(\ref{eq:X_internal}), the relative
error of Eq.~(\ref{eq:differenceQuotient}) is $\mathcal{O}(L^{1/\nu-2-d})$.

We now consider the isotropic XY model on a simple cubic lattice of
size $\Lp\times\Lp\times\Ls$ in three dimensions with periodic boundary
conditions in the parallel directions. The Dirichlet boundary conditions
in perpendicular direction are implemented by open boundary conditions,
which are known to be equivalent at large length scales \cite{Diehl97a,ZhangNhoLandau06},
although alternative implementations are possible \cite{SchultkaManousakis95}.
The Hamiltonian reads \begin{equation}
\mathcal{H}=-\frac{J}{2}\sum_{\langle ij\rangle}\vec{s}_{i}\cdot\vec{s}_{j},\label{eq:Hamiltonian}\end{equation}
where $J>0$ is the ferromagnetic exchange interaction, $\vec{s}_{i}$
are 2-component unit vectors at site $i$, and the sum is restricted
to nearest neighbors on the lattice. The simulations were performed
for several system sizes with fixed aspect ratios $\rho=\Ls/\Lp=1{:}8\,\mathrm{and}\,1{:}16$
using the standard Wolff cluster algorithm \cite{Wolff89}. To calculate
Eq.~(\ref{eq:differenceQuotient}), systems with thicknesses $\Ls'=\Ls\pm1$
at constant $\Lp$ were simulated for every combination of $\Lp$
and $\Ls$. At least $10^{5}$ Monte Carlo sweeps per data point were
performed.

\begin{figure}[t]
\includegraphics[scale=0.6]{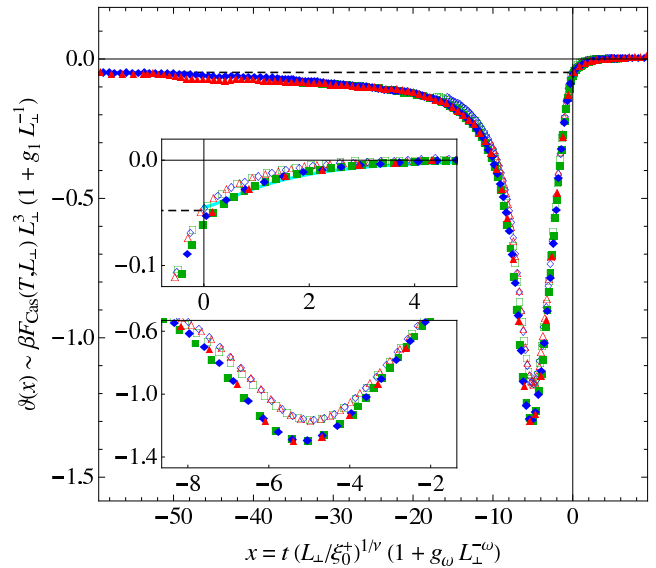}

\caption{\label{fig:X_Casimir}Results for the Casimir force using Eq.~(\ref{eq:ansatz_Casimir}),
for systems with $\Ls=8$ (green squares), $\Ls=12$ (blue diamonds),
and $\Ls=16$ (red triangles), with aspect ratios $\rho=1{:}8$ (open)
and $\rho=1{:}16$ (filled). The statistical error is of the order
of the symbol size. The upper and lower insets are magnifications
around $x=0$ and around the minimum, respectively. Also shown are
the Goldstone amplitude $-\zeta(3)/8\pi$ \cite{LiKardar91} (dashed
line) and the field theoretical result \cite{KrechDietrich92c} (cyan
curve in upper inset). (color online) }
\end{figure}
\begin{figure*}[t]
\includegraphics[scale=0.6]{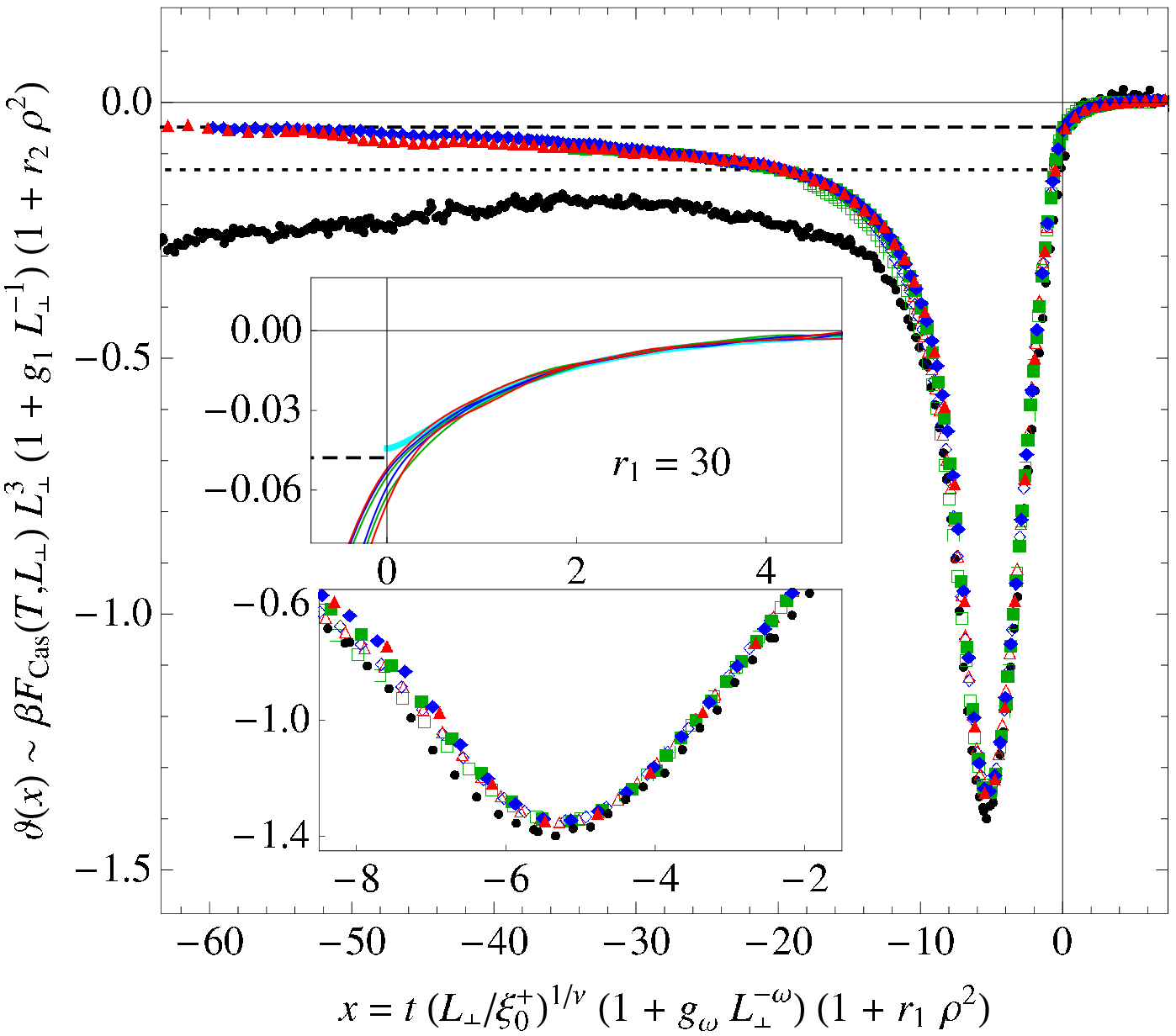}\hfill{}\includegraphics[scale=0.6]{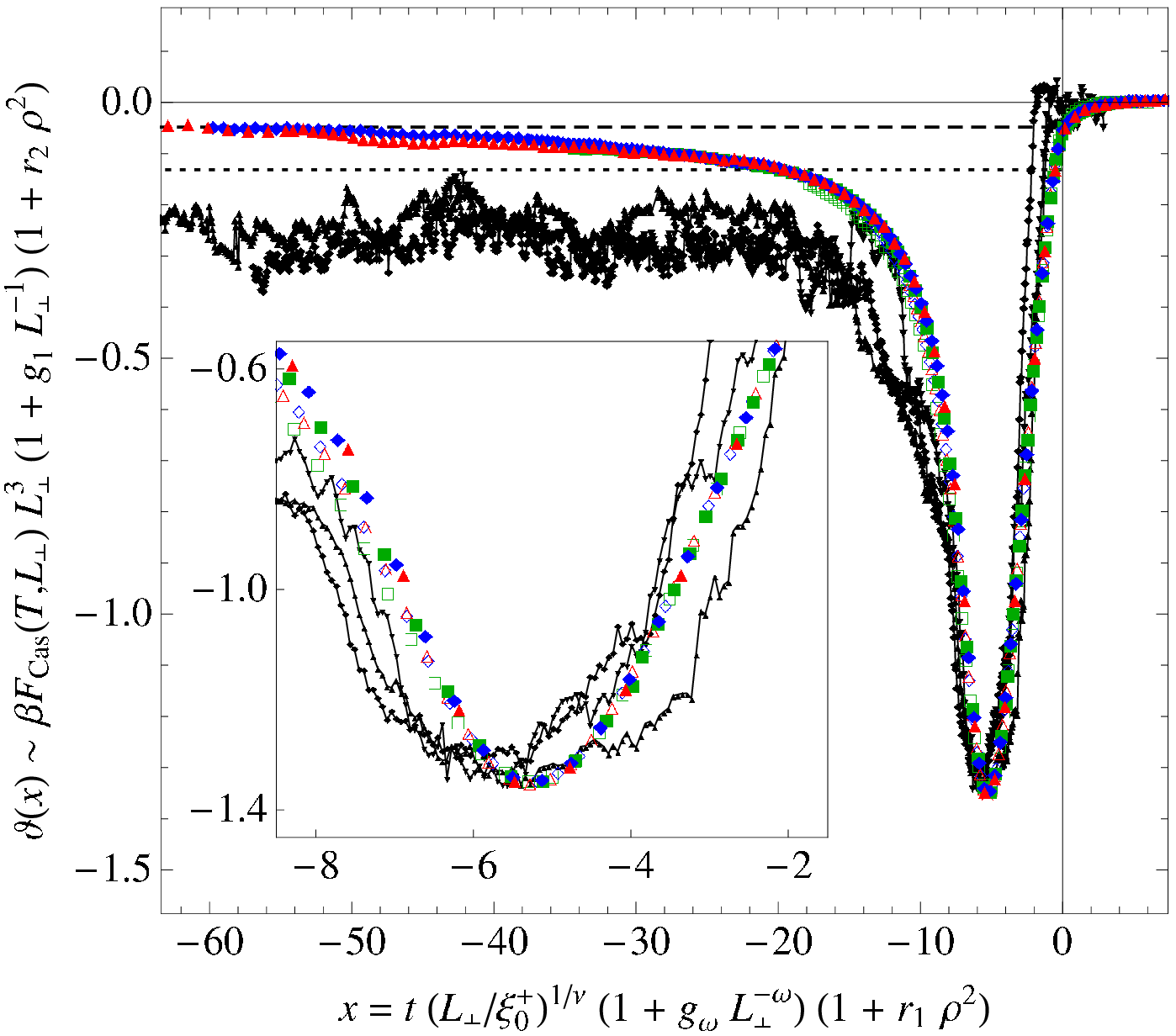}

\caption{\label{fig:X_Casimir2}Universal finite-size scaling function $\Xc(x)$
of the Casimir force for systems with $\Ls=8$ (green squares), $\Ls=12$
(blue diamonds), and $\Ls=16$ (red triangles), with aspect ratios
$\rho=1{:}8$ (open) and $\rho=1{:}16$ (filled). The insets are magnifications
of the respective regions. The results are compared (left) to the
experimental data of Garcia and Chan \cite[Cap.\,1]{GarciaChan99url}
($\bullet$), as well as with the results (right) of Ganshin \emph{et
al.}~\cite{GanshinScheidemantelGarciaChan06} ($\blacktriangle:340\mathring{\mathrm{A}},\,\blacklozenge:285\mathring{\mathrm{A}},\,\blacktriangledown:238\mathring{\mathrm{A}}$;
solid lines are guides to the eye). Also shown are the Goldstone amplitude
$-\zeta(3)/8\pi$ \cite{LiKardar91} (dashed line), the value $-11\zeta(3)/32\pi$
including surface fluctuations proposed in \cite{ZandiRudnickKardar04}
(dotted line), and the field theoretical result \cite{KrechDietrich92c}
(cyan curve in left upper inset). (color online) }
\end{figure*}
To calculate the Casimir forces, Eqs.~(\ref{eq:betaF_Casimir}) and
(\ref{eq:betaF_internal}), it is necessary to have an expression
for the bulk internal energy density $u_{\infty}(T)$. This is achieved
using a combination of direct simulations of a large cubic system
($L=96$) with periodic boundary conditions and results of Cucchieri
\emph{et al.}~\cite{Cucchieri02}. They determined the scaling behavior
of the internal energy and specific heat of the XY model (\ref{eq:Hamiltonian})
in the region $|t|<0.015$, where finite-size effects arise, using
the scaling \emph{ansatz} \begin{eqnarray}
k_{\mathrm{B}}Tu_{\infty}(T) & = & \epsilon_{\mathrm{ns}}+\Tc t\left[C_{ns}+\frac{A^{\pm}}{\alpha}|t|^{-\alpha}\left(\frac{1}{1-\alpha}+\right.\right.\nonumber \\
 &  & \left.\left.+\frac{c_{1}^{\pm}}{1-\alpha+\nu\omega}|t|^{\nu\omega}+\frac{c_{2}^{\pm}}{2-\alpha}t\right)\right].\label{eq:uinf}\end{eqnarray}
The critical indices of the considered XY model are fixed to the values
$\nu=0.672(1)$, $\alpha=-0.017(3)$, $\omega=0.79(2)$, $\Tc/J=2.20183(1)$,
and $\xi_{0}^{+}=0.484(5)$ in the present letter, leading to the
parameters $\epsilon_{\mathrm{ns}}=-0.98841(3)$, $C_{\mathrm{ns}}=22.03$,
$A^{+}=0.3790(8)$, $A^{-}=0.3533(8)$, $c_{1}^{+}=0.015(1)$, $c_{1}^{-}=0.109(2)$,
$c_{2}^{+}=-0.041(3)$, and $c_{2}^{-}=0.211(4)$ \cite{Cucchieri02}.
For $|t|>0.015$ finite-size effects are negligible; note that at
$t=0.015$ the correlation length, Eq.~(\ref{eq:xi}), has the value
$\xi_{\infty}^{+}(0.015)\approx8.1$, which is sufficiently small
with respect to $L=96$. While for periodic cubic systems the scaling
corrections are moderate, systems with broken translational invariance
and aspect ratios $\rho\ll1$ show strong corrections to scaling.
An analysis of usual thermodynamic quantities like the magnetic susceptibility
$\chi(T,\Ls)$ and the Binder cumulant $U(T,\Ls)$ shows that it is
necessary to use a modified scaling variable $x$ (Eq.~(\ref{eq:x}))
with Wegner corrections \cite{Wegner72a} of the form\begin{equation}
x=t\,\left(\frac{\Ls}{\xi_{0}^{+}}\right)^{1/\nu}(1+g_{\omega}\Ls^{-\omega}),\label{eq:x_corr}\end{equation}
while the $y$-direction has rather small corrections for systems
with constant $\rho$. However, for the numerical derivative with
respect to $\Ls$ in (\ref{eq:betaF_internal}) it is necessary to
combine data of systems with $\Ls'=\Ls\pm1$, leading to systems with
different aspect ratio $\rho'\ne\rho$. This and the expected uncertainty
of the numerical derivative itself introduces a scaling correction
of the order $(1+g_{1}\Ls^{-1})$ in the $y$-direction, leading to
the final scaling \emph{ansatz} \begin{equation}
\beta\Fc(T,\Ls)\sim\Ls^{-d}(1+g_{1}\Ls^{-1})^{-1}\Xc(x)\label{eq:ansatz_Casimir}\end{equation}
with $x$ from Eq.~(\ref{eq:x_corr}).

The results for the Casimir force are shown in Figure~\ref{fig:X_Casimir}
for six system sizes with $\Ls\in\{8,12,16\}$, each with aspect ratio
$\rho=1{:}8$ and $\rho=1{:}16$. It should be emphasized that the
only fit parameters are the corrections to scaling amplitudes, $g_{\omega}=2.0(1)$
and $g_{1}=5.5(2)$, which are adjusted \cite{fsscale} until the
numerical data collapse onto a single curve, and that there is no
free factor in neither $x$- nor $y$-direction. We can identify a
slight dependence on the aspect ratio $\rho$ in both directions.
As the corrections due to the finite $\rho$ are known to scale approximately
with $\rho^{2}$ \cite{MonNightingale87}, a full data collapse can
be achieved by adding a factor $(1+r_{1}\rho^{2})$ to the $x$-axis
and a factor $(1+r_{2}\rho^{2})$ to the $y$-axis, with $r_{1}=4(1)$
and $r_{2}=10(1)$. Note that these corrections mainly shift the curves
for $\rho=1{:}8$, while the $\rho=1{:}16$ curves are virtually unchanged
within the error bars. The resulting scaling function is depicted
in Figure~\ref{fig:X_Casimir2}, together with the results of Garcia
and Chan \cite{GarciaChan99url} (left) as well as Ganshin \emph{et
al.}~\cite{GanshinScheidemantelGarciaChan06} (right). In the first
case, only data for the thickest film with $d=423\,\mathring{\mathrm{A}}$
are shown, which are regarded to have the highest quality \cite{Garcia-privcomm},
while the thinner films showed deviations in $y$-direction, presumably
due to surface roughness \cite{GarciaChan99url,GanshinScheidemantelGarciaChan06}.
In order to compare the experimental data quantitatively with the
present results, they are made dimensionless in $x$-direction using
the measured correlation length amplitude $\xi_{0}^{+}=1.432\,\mathring{\mathrm{A}}$
of $^{4}$He at $T_{\lambda}$ \cite{TamAhlers85}, leading to a factor
$(\xi_{0}^{+})^{-1/\nu}=0.586\,\mathring{\mathrm{A}}^{-1/\nu}$. 

We find an excellent agreement within the error bars with both measurements
for $x\gtrsim-8$. The universal amplitude of $\vartheta(x)$ at the
minimum is $\Xc(x_{\mathrm{min}})=-1.35(3)$ at $x_{\mathrm{min}}=-5.3(1)$.
These values agree with the values $x_{\mathrm{min}}=-5.4(1)$ of
Garcia and Chan \cite{GarciaChan99url} and $\Xc(x_{\mathrm{min}})=-1.30(3)$
at $x_{\mathrm{min}}=-5.7(5)$ of Ganshin \emph{et al.}~\cite{GanshinScheidemantelGarciaChan06}.
It turns out that the overall agreement with \cite{GarciaChan99url}
is even better than with \cite{GanshinScheidemantelGarciaChan06},
which might be attributed to the smaller fluctuations in $y$-direction
in the data of \cite{GarciaChan99url}, mainly visible below $x\approx-10$,
and to the five times smaller error estimate in $x$-direction, clearly
visible in the insets and at $x=0$. For $x\lesssim-8$ we see an
enhancement of the measured Casimir force not present in the calculated
scaling function. This onset is weaker in the left figure. A possible
explanation is the occurrence of surface fluctuations below this temperature,
as proposed in \cite{ZandiRudnickKardar04}. At the critical point
($x=0$) we find $\Xc(0)=-0.047(2)\,[-0.059(2)]$ for $\rho=1{:}8\,[1{:}16]$,
which gives $\Xc(0)=-0.062(5)$ for $\rho\rightarrow0$. The resulting
Casimir amplitude $\Delta=-0.031(3)$ agrees well with the estimate
$\Delta=-0.03$ from \cite{Indekeu86a,MonNightingale87}. For $x\gtrsim1$
the results nicely lie on the scaling function calculated by Krech
and Dietrich \cite{KrechDietrich92c} using renormalization group
theory. Here $r_{1}=30$ was used, which again mainly shifts the data
for $\rho=1{:}8$, see upper inset in Figure~\ref{fig:X_Casimir2}(left).
The quality of the used method, especially of the numerical integration,
is demonstrated by the convergence of the calculated scaling function
to the low temperature Goldstone value $\vartheta_{\mathrm{Goldstone}}=-\zeta(3)/8\pi$
\cite{LiKardar91} for $x\rightarrow-\infty$ (dashed line in Figures
\ref{fig:X_Casimir} and \ref{fig:X_Casimir2}).

\begin{figure}[t]
\includegraphics[scale=0.58]{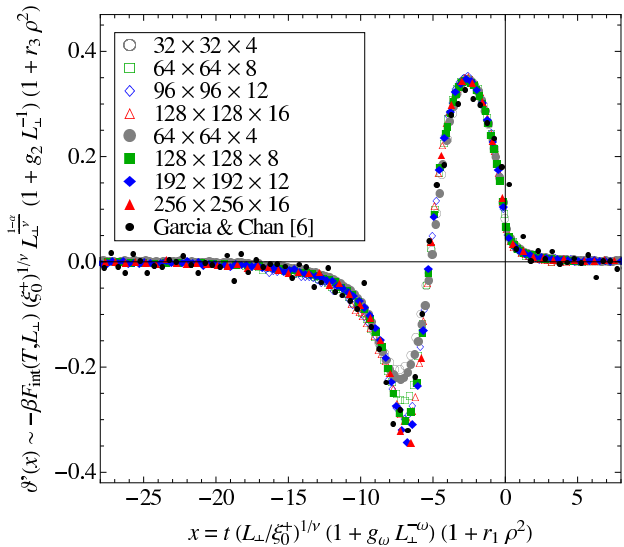}

\caption{\label{fig:X_internal}Universal finite-size scaling function $\Xc'(x)$
of the internal Casimir force, Eq.~(\ref{eq:betaF_internal}), determined
from eight different system sizes with aspect ratios $\rho=1{:}8$
and $1{:}16$. The experimental results are obtained by binning (with
$\Delta x=0.5$) and numerical differentiation of the data from Garcia
and Chan \cite[Cap.\,1]{GarciaChan99url}. (color online) }
\end{figure}
 Figure~\ref{fig:X_internal} shows the results for the scaling function
$\Xc'(x)$ of the internal Casimir force, Eq.~(\ref{eq:betaF_internal}).
The scaling plot contains data from eight system sizes with $\Ls\in\{4,8,12,16\}$,
each with aspect ratio $\rho=1{:}8$ and $\rho=1{:}16$. While the
aspect ratio correction in $y$-direction becomes $r_{3}=r_{2}-r_{1}$,
the scaling correction $g_{2}$ cannot be expressed through $g_{\omega}$
and $g_{1}$, as the corresponding $\Ls$-exponents are different.
The effective calculated value at $\Ls\approx10$ is $g_{2}=1.7$,
which is modified to $g_{2}=2.0(2)$ to get the best data collapse.
The results are compared to a numerical differentiation of the experimental
data of Garcia and Chan \cite[Cap.\,1]{GarciaChan99url}. The results
of Ganshin \emph{et al.}~\cite{GanshinScheidemantelGarciaChan06}
are not shown due to large fluctuations of the numerical derivative.
The data collapse and the agreement with the experimental data is
very convincing, also showing the small influence of statistical errors
in the simulations. Only in the interval $-9\lesssim x\lesssim-5$
we see higher order corrections to scaling, which are believed to
stem from uncertainties in the numerical derivative, Eq.~(\ref{eq:differenceQuotient}).
However, these corrections only have small influence on the integrated
Casimir force. Further work is necessary to clarify this behavior.
Note that at the minimum of $\Xc(x)$ we have $\Xc'(x_{\mathrm{min}})=0$,
which implies that $\lim_{\Ls\rightarrow\infty}\Fi(T_{\mathrm{min}}(\Ls),\Ls)=0$
and\begin{equation}
\left.\frac{\partial u(T,\Ls)}{\partial\Ls}\right|_{T=T_{\mathrm{min}}(\Ls)}\sim u_{\infty}(T_{\mathrm{min}}(\Ls)),\label{eq:umin}\end{equation}
i.e. at this temperature (and large $\Ls$) the change in internal
energy with $\Ls$ equals the corresponding bulk value. The connection
(\ref{eq:umin}) between the minimum of the Casimir force and the
internal energy shows that the shift of the minimum to negative $x$
is a direct consequence of the strong shift in $\Tc(\Ls)$ in systems
with Dirichlet boundary conditions. This shift is also present in
the exactly solvable two dimensional Ising model \cite{BrankovDantchevTonchev00}.

In summary, I determined the universal finite-size scaling function
$\Xc(x)$ of the Casimir force within the XY universality class with
Dirichlet boundary conditions using Monte Carlo simulations. For sufficiently
small aspect ratio $\rho=1{:}16$ the results are in excellent agreement
with the experimental results on $^{4}$He by Garcia and Chan \cite{GarciaChan99url},
and by Ganshin \emph{et al.}~\cite{GanshinScheidemantelGarciaChan06},
as well as with theoretical calculations for $T\ge\Tc$ by Krech and
Dietrich \cite{KrechDietrich92a,KrechDietrich92c}. The universal
function $\Xc(x)$ has a deep minimum at $x_{\mathrm{min}}=-5.3(1)$,
with $\Xc(x_{\mathrm{min}})=-1.35(3)$. The results are in conformity
with the assumption that the order parameter in $^{4}$He asymptotically
obeys Dirichlet boundary conditions. The method proposed in this letter
has also been applied to systems with periodic boundary conditions
\cite{HuchtGrueneberg07}, where a similar good agreement with available
results \cite{DantchevKrech04,DiehlGruenebergShpot06} is obtained.
The application to other boundary conditions as well as to Ising and
Heisenberg models is straightforward.

\paragraph{Note added in proof. }

Recently, Vasilyev \emph{et al.} presented an alternative method to
calculate $\vartheta(x)$ using Monte Carlo simulations~\cite{VasilyevGambassiMaciolekDietrich07}.

\begin{acknowledgments}
I would like to thank Daniel Grüneberg, Hans Werner Diehl, Ralf Meyer,
Daniel Dantchev, and Rafael Garcia for useful discussions and comments. 
\end{acknowledgments}
\bibliographystyle{apsrev}
\bibliography{/users/fred/TeX/bib/Physik.bib}

\end{document}